\documentclass[twocolumn, amsmath, amsfonts, amssymb, trackchanges]{aastex62}
\usepackage{mathtools}
\usepackage{natbib}
\usepackage{bm}

\newcommand{\RB}{Rayleigh-B\'{e}nard }

\newcommand{\approptoinn}[2]{\mathrel{\vcenter{
	\offinterlineskip\halign{\hfil$##$\cr
	#1\propto\cr\noalign{\kern2pt}#1\sim\cr\noalign{\kern-2pt}}}}}

\newcommand{\pro}{\ensuremath{\text{Ro}_{\text{p}}}}
\newcommand{\con}{\ensuremath{\text{Ro}_{\text{c}}}}

\usepackage{color}

\graphicspath{{./}{figs/}{../tex/figs/}}

\received{October 19, 2018}
\revised{January 14, 2019}
\accepted{January 15, 2019}
\submitjournal{ApJ}

\shorttitle{Predicting the Rossby number in convective experiments}
\shortauthors{Anders et al.}

\begin{document}
\defcitealias{anders&brown2017}{AB17}
\newcommand{\AB}{\citetalias{anders&brown2017}}

\title{Predicting the Rossby number in convective experiments}

\correspondingauthor{Evan H. Anders}
\email{evan.anders@colorado.edu}

\author[0000-0002-3433-4733]{Evan H. Anders}
\affil{Dept. Astrophysical \& Planetary Sciences, University of Colorado -- Boulder, Boulder, CO 80309, USA}
\affil{Laboratory for Atmospheric and Space Physics, Boulder, CO 80303, USA}
\author{Cathryn M. Manduca}
\affil{Laboratory for Atmospheric and Space Physics, Boulder, CO 80303, USA}
\author[0000-0001-8935-219X]{Benjamin P. Brown}
\affil{Dept. Astrophysical \& Planetary Sciences, University of Colorado -- Boulder, Boulder, CO 80309, USA}
\affil{Laboratory for Atmospheric and Space Physics, Boulder, CO 80303, USA}
\author[0000-0001-8531-6570]{Jeffrey S. Oishi}
\affiliation{Department of Physics and Astronomy, Bates College, Lewiston, ME 04240, USA}
\author{Geoffrey M. Vasil}
\affiliation{University of Sydney School of Mathematics and Statistics, Sydney, NSW 2006, Australia}

\begin{abstract}
  The Rossby number is a crucial parameter describing the degree of rotational constraint on the convective dynamics in stars and planets. 
However, it is not an input to computational models of convection but must be measured ex post facto. 
Here, we report the discovery of a new quantity, the Predictive Rossby number, which is both tightly correlated with the Rossby number and specified in terms of common inputs to numerical models. 
The Predictive Rossby number can be specified independent of Rayleigh number, allowing suites of numerical solutions to separate the degree of rotational constraint from the strength of the driving of convection.  
We examine the scaling of convective transport in terms of the Nusselt number and the degree of turbulence in terms of the Reynolds number of the flow, 
and we find scaling laws nearly identical to those in nonrotational convection at low
Rossby number when the Predictive Rossby number is held constant.
Finally, we describe the boundary layers as a function of increasing turbulence at constant Rossby number.
\end{abstract}


\section{Introduction}
\label{sec:intro}
Rotation influences the dynamics of convective flows in
stellar and planetary atmospheres.
Many studies on the fundamental nature of
rotating convection in both laboratory and numerical settings
have provided great insight into the properties of convection 
in both the rapidly rotating regime 
and the transition to the rotationally unconstrained regime 
\citep{king&all2009, zhong&all2009, schmitz&tilgner2009, king&all2012, julien&all2012, king&all2013, ecke&niemela2014, stellmach&all2014, cheng&all2015, gastine&all2016}
The scaling behavior of heat transport, the nature of convective flow
structures, and the importance of boundary layer-bulk interactions in driving dynamics are well known.
Yet, we do not know of any simple procedure for predicting the magnitude of vortical flow gradients 
purely from experimental control parameters, such as bulk rotation rate and thermal input.

In the astrophysical context,
many studies of rotating convection have investigated questions inspired by the solar dynamo
\citep{glatzmaier&gilman1982, busse2002, brown&all2008,
brown&all2010, brown&all2011, augustson&all2012, guerrero&all2013, kapyla&all2014}.
Even when these simulations nominally rotate at the solar rate,
they frequently produce distinctly different behaviors than the true Sun,
such as anti-solar differential rotation profiles  \citep{gastine&all2014, brun&all2017}.
It seems that these differences occur because the simulations produce less rotationally 
constrained states than the Sun. 
The influence of rotation results from the local 
shear gradients, and these are not direct input parameters.
Recent simulations predict significant rotational influence in the deep solar interior, 
which can drastically affect flows throughout the solar convection zone 
\citep{featherstone&hindman2016, greer&all2016}. 
In the planetary context, the balance between magnetic
and rotational forces likely leads to the observed differences between ice
giant and gas giant dynamos in our solar system \citep{soderlund&all2015}.
The work of \cite{aurnou&king2017} demonstrates the importance of studying a dynamical regime
with the proper balance between Lorentz, Coriolis, and inertial forces when modeling
astrophysical objects such as planetary dynamos.

In short, simulations must achieve the proper rotational balance if they are to explain 
the behavior of astrophysical objects. 
In Boussinesq studies, rotational constraint is often measured by comparing
dynamical and thermal boundary layers or deviation in heat transport from the non-rotating
state \citep{king&all2012, julien&all2012, king&all2013}. 
Such measurements are not available for astrophysical objects, where
the degree of rotational influence is best assessed by the ratio between 
nonlinear advection magnitude and the linear Coriolis accelerations. 
The \textit{Rossby number} is the standard measure of this ratio, 
\begin{equation}
\text{Ro} \ \equiv \ \frac{| \nabla \times \boldsymbol{u} | }{2 |\bm{\Omega}|} \ 
\sim \ \frac{| (\nabla \times \boldsymbol{u}) \times \boldsymbol{u}  | }{|2 \bm{\Omega} \times \boldsymbol{u}|},
\label{eqn:rossby-def}
\end{equation}
where $\bm{\Omega}$ denotes the bulk rotation vector. 
Many proxies for the dynamical Rossby number exist that are based solely on input parameters, most notably the \textit{convective} Rossby number. 
However, all proxies produce imperfect predictions for the true dynamically relevant quantity.
\begin{quote}
\emph{In this letter, we demonstrate an emperical method of predicting the output Rossby number
of convection in a simple stratified system.}
\end{quote}
In \cite{anders&brown2017} (hereafter \AB), we studied non-rotating compressible convection without magnetic fields in polytropic atmospheres. 
In this work, we extend \AB$\,$ to rotationally-influenced, $f$-plane
atmospheres 
\cite[e.g.,][]{brummell&all1996, brummell&all1998, calkins&all2015a}. 
We determine how the input parameters we studied previously, which controlled the Mach and
Reynolds numbers of the evolved flows, couple with the Taylor number \citep[Ta,][]{julien&all1996}, which sets the magnitude of the rotational vector. 

In section  \ref{sec:experiment}, we describe our experiment and paths through parameter space. 
In section \ref{sec:results}, we present the results of our experiments and in section \ref{sec:discussion} we offer concluding remarks.

\section{Experiment} 
\label{sec:experiment}
We study fully compressible, stratified 
convection under precisely the same atmospheric model
as in \AB, but here
we have included rotation. We study polytropic atmospheres
with $n_\rho = 3$ density scale heights and a superadiabatic
excess of $\epsilon = 10^{-4}$ such that flows are at low Mach number.
We study a domain in which the
gravity, $\bm{g} = -g\hat{z}$, and rotational vector, $\bm{\Omega} = \Omega \hat{z}$, 
are antiparallel \citep[as in e.g.,][]{julien&all1996, brummell&all1996}.

We evolve the velocity ($\bm{u}$), temperature ($T$), 
and log density ($\ln\rho$) according to the Fully Compressible Navier-Stokes equations
in the same form presented in \AB, with the
addition of the Coriolis term, \mbox{$2\bm{\Omega}\times\bm{u}$}, to the left-hand side
of the momentum equation. 
We impose impenetrable, stress-free, fixed-temperature boundary conditions at the top and bottom of the domain.

We specify the kinematic viscosity ($\nu$), thermal diffusivity ($\chi$), and strength of
rotation ($\Omega$) at the top of the domain by choosing the Rayleigh number 
(Ra), Prandtl number (Pr), and Taylor number (Ta),
\begin{equation}
    \text{Ra} = \frac{g L_z^3 \Delta S / c_P}{\nu \chi}, \,\,\,
    \text{Pr} = \frac{\nu}{\chi}, \,\,\,
    \text{Ta} = \left(\frac{2 \Omega L_z^2}{\nu}\right)^2,
	\label{eqn:input_parameters}
\end{equation}
where $L_z$ is the depth of the domain as defined in \AB, 
$\Delta S \propto \epsilon n_\rho$ is the specific entropy difference between
the top and bottom of the atmosphere, and the specific heat at constant pressure is $c_P = \gamma/(\gamma-1)$
with $\gamma = 5/3$.
Throughout this work we set Pr = 1. The Taylor number relates to the often-quoted
Ekman number by the equality $\text{Ek} \equiv \text{Ta}^{-1/2}$.

Due to stratification, Ra and Ta both grow with depth as (Ra,Ta)$\, \propto \rho^2$ (see \AB).
We nondimensionalize our atmospheres at the top of the domain, and so all values of Ra and
Ta quoted in this work are the minimal value of Ra and Ta in the domain at $z = L_z$.
For direct comparison to Boussinesq studies, past work has found that the value of Ra 
at the atmospheric midplane ($z = L_z/2$) varies minimally with increasing stratification 
\citep{unnoetall1960}. For the atmospheres presented in this work, 
midplane Ra and Ta values
are larger than reported top-of-atmosphere values by a factor of $\sim$70, and values at the
bottom of the atmosphere are larger by $\sim$400.

When Ta is large, the wavenumber of convective onset increases according to $k_{\text{crit}} \propto \text{Ta}^{1/6}$
\citep{Chandrasekhar,calkins&all2015a}.
We study horizontally-periodic, 3D Cartesian domains with extents of
$x, y = [0, 4(2\pi/k_{\text{crit}})]$ and $z = [0, L_z]$. At large values of Ta, these
domains are tall and skinny, as in \cite{stellmach&all2014}.
We evolve our simulations using the Dedalus\footnote{\url{http://dedalus-project.org/}} 
pseudospectral framework, and our numerical methods are identical to those presented
in \AB. The supplemental materials of this paper include a \texttt{.tar} file which
contains the code used to perform the simulations in this work,
and this tarball is also published online in a Zenodo repository \citep{supp_andersetall2019}.

\begin{figure*}[t!]
    \includegraphics[width=\textwidth]{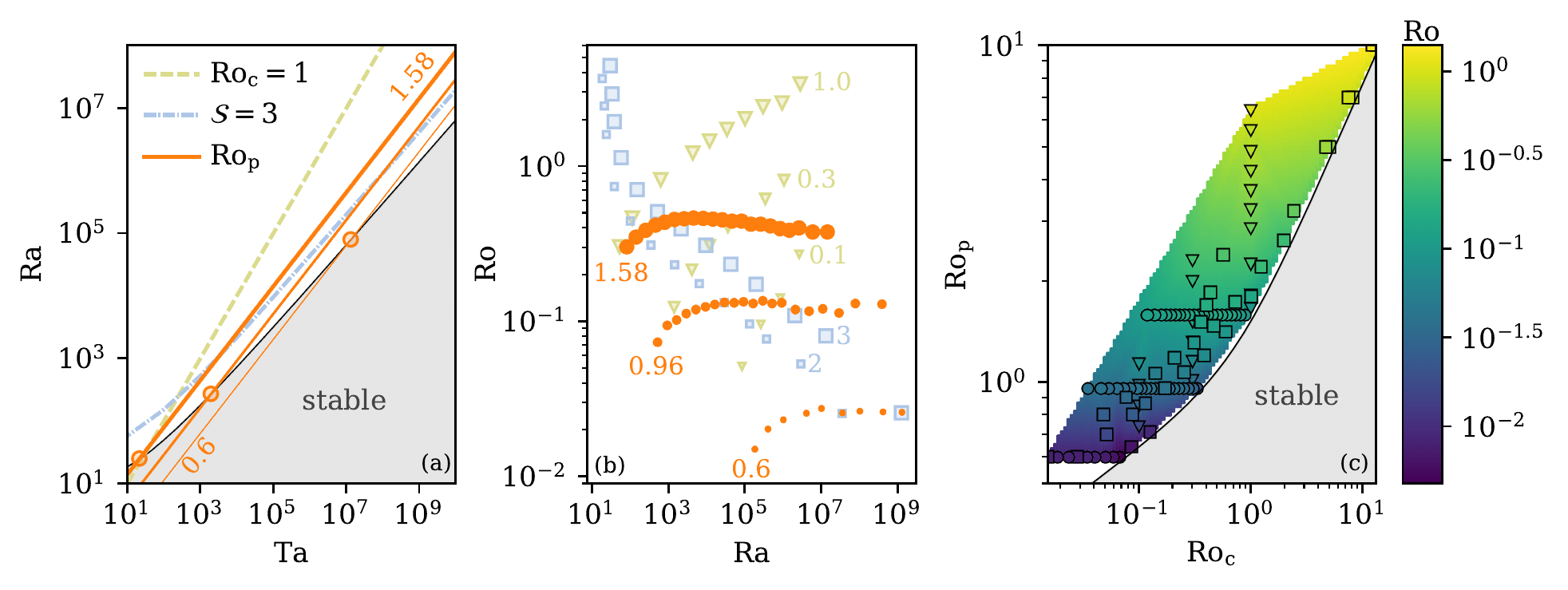}
    \caption{(a) The critical Rayleigh number, as a function of the Taylor number, 
    is plotted as a solid black line. The grey shaded region is subcritical, and rotation
    supresses convection there. Paths of constant Convective Rossby number
    ($\con$, green dashed line), constant supercriticality ($\mathcal{S}$, blue dash-dot line), and 
    constant Predictive Rossby number (\pro, orange solid lines) are shown. 
	From thickest to thinnest, paths with $\pro = [1.58, 0.96, 0.6]$ are plotted,
	and the value of
	($\text{Ta}_{\text{crit}}, \text{Ra}_{\text{crit}}$) for each path is denoted by a
	circular marker (see Table \ref{table:ra_ta_bounds}).
    (b) Evolved Ro is plotted vs. Ra along paths of \pro = [1.58, 0.96, 0.6] for [big, medium, small] orange circles.
    For comparison, paths of constant $\mathcal{S}$ (blue squares, $\mathcal{S} = [3,2]$ 
    for [big, small] squares)
    and constant $\con$ (green triangles, $\con$ = [1, 0.3, 0.1] for [big, medium, small] triangles) are shown.
    (c) The evolved value of Ro is shown as a function of $\pro$ and $\con$. 
	Each of the experiments in (b) is outlined by a black (circle, triangle, square)
	for points along constant (\pro, \con, $\mathcal{S}$) paths.
	The color inside of the marker represents the exact measured Ro of that experiment, 
	while the colormap outside of markers is a linear interpolation
	of the data set. 
    \label{fig:parameter_space} }
\end{figure*}

The critical value of Ra at which rapidly rotating convection onsets also depends on Ta (see the black line in figure \ref{fig:parameter_space}a),
roughly according to $\text{Ra}_{\text{crit}} \sim \text{Ta}^{2/3}$ \citep{Chandrasekhar,calkins&all2015a}.
Even taking account of linear theory, the dependence of the evolved nonlinear fluid 
flows on the input parameters makes predicting the rotational constraint very challenging. 
We will explore three paths through Ra-Ta space:
\begin{equation}
    \text{Ra} = 
    \begin{cases}
    \mathcal{S}\,\text{Ra}_\text{crit}(\text{Ta}), & (\text{I})\\
    (\con)^2 \, \text{Pr}\, \text{Ta}, & (\text{II}) \\
    (\pro)^2\, \text{Pr}^{1/2}\, \text{Ta}^{3/4} & (\text{III}).
    \end{cases}
    \label{eqn:paths}
\end{equation}
Paths on constraint I are at constant supercriticality, 
$\mathcal{S} \equiv \text{Ra}/\text{Ra}_{\text{crit}}(\text{Ta})$
(blue dash-dot line in figure \ref{fig:parameter_space}a).
Paths on constraint II (green dashed line in figure \ref{fig:parameter_space}a)
are at a constant value of the classic
\textit{convective} Rossby number, 
\begin{equation}
\con \ = \   \sqrt{ \frac{\text{Ra}}{\text{Pr}\, \text{Ta} } } \ 
= \  \frac{1}{2 \Omega } \sqrt{\frac{g \, \Delta  S}{c_{p} L_{z}}},
\label{eqn:roc_defn}
\end{equation}
which has provided \citep[e.g., ][]{julien&all1996, brummell&all1996} 
a common proxy for the degree of rotational constraint.
This parameter measures the importance of buoyancy relative to rotation without 
involving dissipation.  
Paths on constraint
III (e.g., orange solid lines in figure \ref{fig:parameter_space}a) 
set constant a ratio which we call the ``Predictive Rossby number,'' 
\begin{equation}
\pro = \sqrt{\frac{\text{Ra}}{\text{Pr}^{1/2}\,\text{Ta}^{3/4}}} \ = \    
\frac{1}{(2 \Omega)^{3/4}} \sqrt{\frac{g \, \Delta  S}{c_{p} \chi^{1/2}}}
\label{eqn:rop_defn}
\end{equation}
Unlike paths through parameter space which hold \con$\,$ constant,
paths with constant \pro$\,$ 
feel changes in diffusivities but not the depth of the domain.
To our knowledge, these paths have not been reported in the literature, 
although the importance of $\text{Ra}/\text{Ta}^{3/4} = \text{Ra Ek}^{3/2}$
has been independently found by \cite{king&all2012} using a boundary layer
analysis. We compare our results to their theory in Section \ref{sec:discussion}. 

In this work, we primarily study three values of \pro. These values are shown in
Fig.~\ref{fig:parameter_space}a and Table \ref{table:ra_ta_bounds}. Table \ref{table:ra_ta_bounds}
lists the values of (Ra$_{\text{crit}}$, Ta$_{\text{crit}}$) for each value of \pro, and also
the maximum value of (Ra, Ta) studied in this work for each path. We additionally walked
two pathways at constant supercriticality (constraint I, $\mathcal{S} = \{2, 3\}$) and
three pathways at constant convective Rossby number (constraint II, $\con = \{1, 0.3, 0.1\}$).
Full details on all cases are provided in Appendix \ref{appendix:table} and the supplemental
materials.

\begin{deluxetable}{c c c}
\caption{Parameter space 
\label{table:ra_ta_bounds}
}
\tablehead{\colhead{\pro} & \colhead{(Ra$_{\text{crit}}$, Ta$_{\text{crit}}$)} & \colhead{(Ra$_{\text{max}}$, Ta$_{\text{max}}$)}}
\startdata
0.60 & ($10^{4.88}, 10^{7.10}$) & ($10^{9.09}, 10^{12.72}$)\\
0.96 & ($10^{2.44}, 10^{3.30}$) & ($10^{8.58}, 10^{11.49}$)\\
1.58 & ($10^{1.39}, 10^{1.33}$) & ($10^{7.14}, 10^{8.99}$)\\
\enddata
\tablecomments{
Values of the critical Ra and Ta for each \pro$\,$ track are reported, as well as the
maximal values of Ra and Ta studied on each track. All values reported are for the top of
the atmosphere. A fuller set of simulations are reported in Table \ref{table:simulation_info}
with midplane Ra and Ta values as well.
}
\end{deluxetable}

\section{Results}
\label{sec:results}
In our stratified domains, for $\text{Ta} \geq 10^5$, 
a best-fit to results from a linear stability
analysis provides $\text{Ra}_{\text{crit}}(\text{Ta}) = 1.459\text{Ta}^{2/3}$ 
and $k_{\text{crit}}(\text{Ta}) = 0.414\text{Ta}^{1/6}$ for direct onset of convection.
In figure~\ref{fig:parameter_space}a, the value of Ra$_{\text{crit}}(\text{Ta})$
is shown. Sample paths for
each criterion in equation~\ref{eqn:paths} through
this parameter space are also shown.
In this work, we often find it instructive to use one critical Ra for an entire \pro$\,$ path.
This $\text{Ra}_\text{crit}$ is determined by the intersection of the onset curve and \pro$\,$ path
(indicated by the orange circles in figure~\ref{fig:parameter_space}a, and quoted in Table \ref{table:ra_ta_bounds}). 
In the high Ta regime, we find that $\text{Ra}_{\text{crit}} = 18.5\pro^{-16}$.

In figure \ref{fig:parameter_space}b, we display the evolution of Ro
with increasing Ra along various paths through parameter space.
We find that Ro increases on constant \con$\,$paths, decreases on constant $\mathcal{S}$
paths, and remains roughly constant along constant \pro$\,$ paths.
In figure \ref{fig:parameter_space}c, the value of Ro is shown simultaneously as
a function of \pro$\,$and \con$\,$for all experiments conducted in this study.
We find a general power-law of the form \mbox{$\text{Ro} = \mathcal{C} \con^{\alpha}\,\pro^{\beta}$}.
In the rotationally-dominated regime where $\text{Ro} < 0.2$ and 
$\text{Re}_{\perp} > 5$ (see Eqn. \ref{eqn:re_defn}),
we find $\alpha = -0.02$, and $\text{Ro}$ can be said to be a function
of $\pro$ alone. Under this assumption, we report a scaling of $\text{Ro} = (0.148 \pm 0.003) \pro^{3.34 \pm 0.07}$.
In the less rotationally dominated regime of $\text{Ro} > 0.2$ and $\text{Re}_{\perp} > 5$, 
we find $\{C, \alpha, \beta\} = \{0.2, -0.19, 1.5\}$.

\begin{figure*}[t]
    \includegraphics[width=\textwidth]{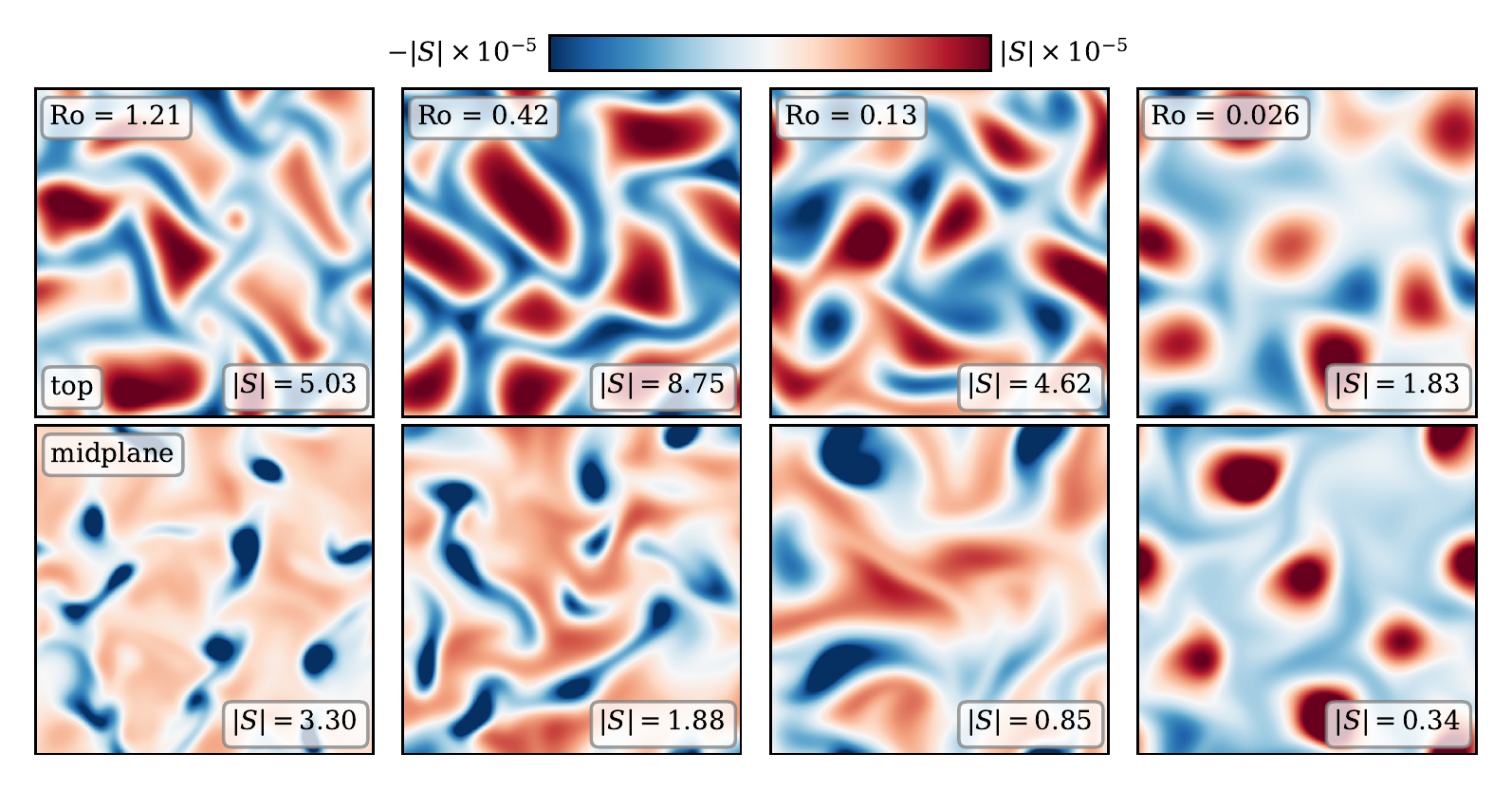}
    \caption{ Horizontal slices of the evolved entropy deviation from the mean
	at $z = 0.95L_z$ (top row) and $z = 0.5L_z$ (bottom row) are shown for select simulations. 
	All runs displayed here have an evolved volume-averaged $\text{Re}_\perp \approx 32$. 
    As Ro decreases from O(1) on the left to O(0.03) on the right, and thus the rotational
    constraint on the flow increases, significant changes in flow morphology are observed.
    As Ro decreases, Coriolis forces more effectively
    deflect the convective flows, and the classic granular convective pattern gives way
    to vortical columns that are quasi-two-dimensional.
    \label{fig:dynamics_plot} }
\end{figure*}

\begin{figure}[t!]
    \includegraphics{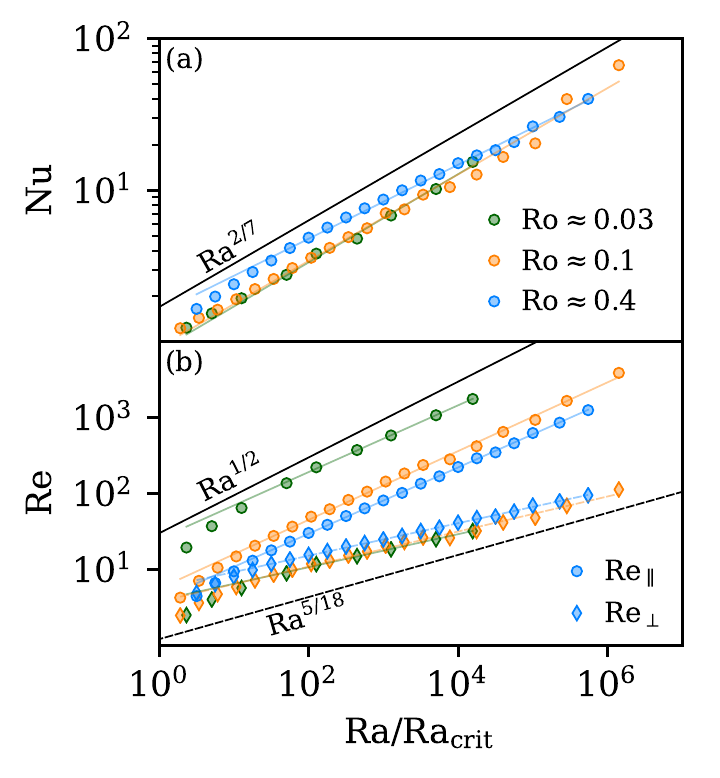}
    \caption{
	Scaling laws for paths at $\pro = 1.58$ ($\text{Ro} \approx 0.4$),
    $\pro = 0.96$ ($\text{Ro} \approx 0.1$), and $\pro = 0.6$ ($\text{Ro} \approx 0.03$) are shown. 
    Numbers are plotted vs. Ra/Ra$_{\text{crit}}$, where Ra$_{\text{crit}}$ is given in Table \ref{table:ra_ta_bounds}.
	(a) Nu, as defined in \AB, is shown.
    (b) $\text{Re}_\parallel$ and $\text{Re}_{\perp}$, as defined in equation \ref{eqn:re_defn},
	are shown. All values of $\pro$ trace out similar Nu and $\text{Re}_{\perp}$ tracks,
	whereas $\text{Re}_\parallel$ tracks shift upwards as Ro decreases.
    \label{fig:nu_and_re} }
\end{figure}

\begin{figure*}[ht!]
    \includegraphics[width=\textwidth]{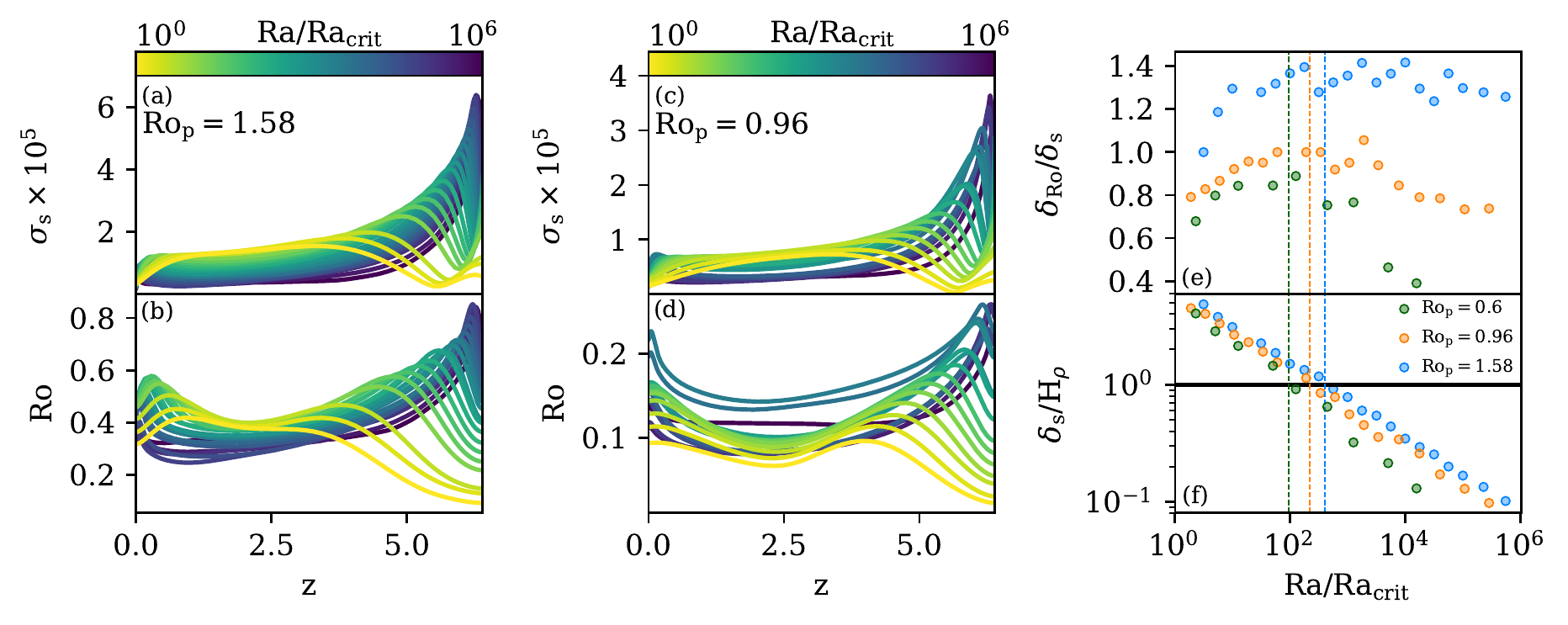}
    \caption{Horizontally-averaged profiles of the standard deviation of entropy ($\sigma_{\text{s}}$, a) and Rossby number (Ro, b) 
    are shown vs. height for $\pro = 1.58$ ($\text{Ro} \approx 0.4$). 
    Similar profiles are shown in (c) and (d) for $\pro = 0.96$ ($\text{Ro} \approx 0.1$). The color of the profiles
    denotes the value of Ra/Ra$_{\text{crit}}$, where Ra$_\text{crit}$ is given in Table \ref{table:ra_ta_bounds}.
    (e) The ratio of the thicknesses of the dynamical boundary layers ($\delta_{\text{Ro}}$) and 
    thermal boundary layers ($\delta_{\text{s}}$) is shown vs. $\text{Ra}/\text{Ra}_{\text{crit}}$ for fixed \pro.
	(f) $\delta_{\text{s}}$ is plotted vs. $\text{Ra}/\text{Ra}_{\text{crit}}$ in units of the density
	scale height at the top of the atmosphere ($H_\rho$). Vertical lines denote when $\delta_\text{s}/H_\rho = 1$
	for each value of \pro.
    \label{fig:profiles_and_bls} }
\end{figure*}

In figure~\ref{fig:dynamics_plot}, sample snapshots of the evolved entropy field 
in the $x$-$y$ plane near the top and at the middle of the domain are shown. 
In the left column, flows are at $\text{Ro} \sim 1$ and resemble the classic granular structure 
of nonrotating convection (see e.g., figure~2 in \AB), where strong narrow downflow lanes 
punctuate broad upwellings. The narrow downflows at the top organize themselves into intense
coherent structures at the midplane, and at the midplane the downflows have much stronger
entropy fluctuations than the broad and slower upflows. 

Ro decreases from left to right into the rotationally constrained 
regime. As Ro decreases, the narrow downflow lanes begin to disappear
and the flows at midplane become more symmetric. In the rotationally constrained regime ($\text{Ro} \sim 0.03$), 
the convective structures are distinctly different. 
Here we observe dynamically persistent, warm upflow columns 
surrounded by bulk weak downflow regions. At the midplane, the upflow columns have substantially
higher entropy perturbations than the surrounding weak downflows which sheathe them, 
and the locations of the columns
are tightly correlated with their positions at the top of the domain. These quasi-two-dimensional 
dynamics are similar to those seen in rapidly 
rotating \RB convection \citep[e.g.,][]{stellmach&all2014}. 
The select cases displayed in figure~\ref{fig:dynamics_plot} each have an evolved volume-averaged
$\text{Re}_{\perp} \approx 32$ (defined below in equation~\ref{eqn:re_defn}).

We measure the Nusselt number (Nu), which quantifies heat transport in a convective
solution, as defined in \AB.
In figure~\ref{fig:nu_and_re}a, we plot Nu as a function
of $\text{Ra}/\text{Ra}_\text{crit}$ at fixed \pro. We find that $\text{Nu} \propto \{\text{Ra}^{0.29 \pm 0.01}, \text{Ra}^{0.29 \pm 0.01}, \text{Ra}^{0.24}\}$
for $\pro = \{0.6, 0.957, 1.58\}$. In the regime of $\text{Ro} \lesssim 0.1$, these scaling laws are indistinguishable
from a classic Ra$^{2/7}$ power law scaling,
which is observed in nonrotating \RB and stratified convection \citep[][\AB]{ahlers&all2009}.
Our results seem consistent with the stress-free, rotating \RB convection results of
\cite{schmitz&tilgner2009}, whose re-arranged Eqn.~7 returns a best-fit of
$\text{Nu} \propto \text{Ra}^{0.26}$ at fixed \pro$\,$. Their work primarily spans the transition
regime between rotationally constrained and unconstrained convection, and so it is perhaps not
surprising that their power law is a blend of our rotationally-constrained $\text{Ra}^{2/7}$ power
law and the fairly rotationally unconstrained $\text{Ra}^{0.24}$ at $\pro = 1.58$.

Flows are distinctly different parallel to and perpendicular
from the rotation vector, which aligns with gravity and stratification.
We measure two forms of the RMS Reynolds number,
\begin{equation}
\text{Re}_{\parallel} = \frac{|\bm{u}| L_z}{\nu}, \qquad
\text{Re}_{\perp}     = \frac{|\bm{u}|}{\nu}\frac{2\pi}{k_{\text{crit}}},
\label{eqn:re_defn}
\end{equation}
where the length scale in $\text{Re}_{\perp}$ is the wavelength of convective onset, and is
related to the horizontal extent of our domain (see section \ref{sec:experiment}).
From our work in \AB, we expect the RMS velocity to scale as $|u| \propto \sqrt{\Delta S}$.
By definition, $\nu \propto \sqrt{\text{Ra}/(\text{Pr }\Delta S)}$, and $L_z$ is a constant
set by the stratification while $k_{\text{crit}} \propto \text{Ta}^{1/6}$. Along paths of
constant \pro, we thus expect $\text{Re}_{\parallel} \propto \text{Ra}^{1/2}$ and
$\text{Re}_{\perp} \propto \text{Ra}^{5/18}$ when $\text{Pr}$ is held constant.

In figure \ref{fig:nu_and_re}b, we plot $\text{Re}_{\parallel}$ and $\text{Re}_{\perp}$
as a function of Ra$/\text{Ra}_{\text{crit}}$ at fixed \pro. We find that $\text{Re}_\parallel \propto \{\text{Ra}^{0.44 \pm 0.01}, \text{Ra}^{0.45 \pm 0.01}, \text{Ra}^{0.44}\}$
and $\text{Re}_\perp \propto \{\text{Ra}^{0.22 \pm 0.01}, \\ \text{Ra}^{0.23 \pm 0.01}, \text{Ra}^{0.21}\}$ for
$\pro = \{0.6, 0.957, 1.58\}$. These scalings are similar to but slightly weaker than our
predictions in all cases. However, the scaling of $\text{Re}_{\parallel} \propto \text{Ra}^{0.45}$, 
is once again a power law observed frequently in nonrotating convection \citep[][\AB]{ahlers&all2009}.
We also observe that $\text{Re}_{\perp}$ collapses for each $\pro\,$ track,
while $\text{Re}_{\parallel}$ experiences an offset
to larger values as $\pro$ shrinks. The offset in $\text{Re}_{\parallel}$ is unsurprising, 
because more rotationally constrained flows result in smaller boundary layers relative to the 
vertical extent of our stratified domain. The horizontal extent of our domain scales with the
strength of rotation, and so regardless of \pro, flows perpendicular
to the rotational and buoyant direction are comparably turbulent at the same 
$\text{Ra}/\text{Ra}_\text{crit}$.
We find $\text{Re}_{\perp}$ and $\text{Re}_{\parallel}$ are, respectively, good proxies for
the horizontal and perpendicular resolution required to resolve an experiment.

Figure \ref{fig:profiles_and_bls} shows time- and horizontally-averaged profiles of
Ro and the standard deviation of the entropy, $\sigma_{\text{s}}$.
Figures \ref{fig:profiles_and_bls}a\&b show these profiles for $\pro=1.58$ ($\text{Ro} \approx 0.4$), while
Figures \ref{fig:profiles_and_bls}c\&d show these profiles for $\pro=0.96$ ($\text{Ro} \approx 0.1$). The transition
in profile behavior from low Ra (yellow) to high Ra (purple) is denoted by the color of the
profile.
As Ra increases at a constant value of
\pro, both the thermal ($\sigma_{\text{s}}$) and dynamical (Ro) boundary layers become thinner. 
We measure the
thickness of the thermal boundary layer ($\delta_{\text{s}}$) at the top of the domain by 
finding the location of the first maxima of $\sigma_{\text{s}}$ away from the boundary.
We measure
the thickness of the Ro boundary layer ($\delta_{\text{Ro}}$) 
in the same manner.
In figure \ref{fig:profiles_and_bls}e, we plot $\delta_{\text{Ro}}/\delta_{\text{s}}$, the ratio
of the sizes of these two boundary layers. As anticipated, the dynamical boundary layer ($\delta_{\text{Ro}}$)
becomes relatively thinner with respect to the thermal boundary layer ($\delta_{\text{s}}$)
as Ro and \pro$\,$ decrease. 
However, the precise scaling of this boundary layer ratio with \pro$\,$ and Ra is unclear, 
and we cannot immediately compare these ratios to similar measures from the \RB convection
literature, such as Fig. 5 of \cite{king&all2013}. They measure the dynamical boundary layer
thickness as the peak location of the horizontal velocities, but our horizontal velocities
are subject to stress-free boundary conditions, and we find that the maxima of horizontal 
velocities occur precisely at the boundaries.
In figure \ref{fig:profiles_and_bls}f, we plot $\delta_\text{s}$ in units of the density
scale height at the top of the atmosphere, and we plot vertical lines when this crosses 1.
We find no systematic change in behavior when $\delta_{\text{s}}$ is smaller than the
local density scale height.

\section{Discussion}
\label{sec:discussion}
In this letter, we studied low-Mach-number, stratified, compressible convection 
under the influence of rotation.
We examined three paths through Ra-Ta space, and showed that the newly-defined 
Predictive Rossby number, $\pro = \text{Ra}/(\text{Pr }^{1/2}\text{Ta}^{3/4})$, determines the value of
the evolved Rossby number. 

Shockingly, along these constant \pro$\,$ pathways, particularly when $\text{Ro} \lesssim 0.1$, 
we find $\text{Nu} \propto \text{Ra}^{2/7}$ and $\text{Re}_\parallel \propto \text{Ra}^{0.45}$.
These scalings are indistinguishable from the scalings of
Re and Nu with Ra in non-rotating Boussinesq convection \citep{ahlers&all2009}.  
\citet{julien&all2012} theorized that in the rapidly rotating
asymptotic limit, $(\text{Nu} - 1) \propto (\text{Ra}^{3/2}/\text{Ta}) = (\text{Ra}/\text{Ra}_{\text{crit}}(\text{Ta}))^{3/2}$. Thus,
at fixed \text{Ta}, a very sharp $\text{Ra}^{3/2}$ scaling law is expected. 
At a fixed Ta = $10^{14}$, \citet{stellmach&all2014} found that the Ra$^{3/2}$ scaling described
the results of stress-free DNS in Boussinesq cylinders very well.  \citet{gastine&all2016} 
studied Boussinesq convection in spherical shells with no-slip boundaries, and also found 
good agreement with the theory of \citet{julien&all2012} for various Ra at $\text{Ta} \geq 10^{10}$.

Here, when we run simulations at fixed \pro, the value of Ta is coupled to the value of Ra, and both increase simultaneously.
Recasting the scaling of 
\citet{julien&all2012} into this perspective, we find 
$(\text{Nu} - 1) \propto \text{Ra}^{3/2}/\text{Ta} = \pro^{8/3}\text{Ra}^{1/6} \propto (\text{Ra}/\text{Ra}_{\text{crit}})^{1/6}$, 
where in this final result we use the $\text{Ra}_{\text{crit}}$ value of the whole $\pro$ path, such as those
specified in Table \ref{table:ra_ta_bounds}.
This $\text{Ra}^{1/6}$ scaling is much weaker than the $\text{Ra}^{2/7}$ law we find here. 
We leave it to future work to explain this discrepancy between
Boussinesq theory and our observed Nu vs.~Ra scaling.

In this work, we experimentally arrived at the 
$\text{Ra}/\text{Ta}^{3/4} = \text{Ra}\,\text{Ek}^{3/2}$
scaling in \pro, but this relationship was independently
discovered by \cite{king&all2012}. Arguing that the thermal boundary layers should
scale as $\delta_{S} \propto \text{Ra}^{-1/3}$ and rotational Ekman boundary layers
should scale as $\delta_{\text{Ro}} \propto \text{Ta}^{-1/4} = \text{Ek}^{1/2}$, they
expect these boundary layers to be equal in size when $\text{Ra}/\text{Ta}^{3/4} \sim 1$.
They demonstrate that when $2 \lesssim\text{Ra}/\text{Ta}^{3/4} \lesssim 20$ flows are in the
transitional regime, and for $\text{Ra}/\text{Ta}^{3/4} \lesssim 2$, flows are rotationally
constrained. We remind the reader that Boussinesq values of $\text{Ra}$ and $\text{Ta}$ are 
not the same as their
values in our stratified domains here, as diffusivities change with depth (see section \ref{sec:experiment}). 
Taking into account this change with depth,
our simulations fall in \cite{king&all2012}'s rotationally constrained
($\pro = 0.6$) and near-constrained transitional regime ($\pro = \{0.957, 1.58\}$).
The measured values of Ro in Fig.~\ref{fig:parameter_space}b and the observed dynamics
in Fig.~\ref{fig:dynamics_plot} agree with this interpretation.

We note briefly that the scaling $\text{Ra} \propto \text{Ta}^{3/4}$ is very similar to
another theorized boundary between fully rotationally constrained convection and 
partially constrained convection predicted in Boussinesq theory, of 
$\text{Ra} \propto \text{Ta}^{4/5}$ \citep{julien&all2012, gastine&all2016}. This
Ta$^{4/5}$ scaling also arises through arguments of geostrophic balance in the boundary layers,
and is a steeper scaling than the Ta$^{3/4}$ scaling present in \pro.
This suggests that at sufficiently low \pro, a suite of simulations across many orders
of magnitude of Ra will not only have the same volume-averaged value of Ro 
(as in Fig.~\ref{fig:parameter_space}b), but will
also maintain proper force balances within the boundary layers.

Our results suggest that by choosing the desired value of \pro, experimenters
can select the degree of rotational constraint present in their simulations. 
We find that $\text{Ro} \propto \text{Ro}_\text{p}^{3.34 \pm 0.07}$, which is within
2$\sigma$ of the estimate in \cite{king&all2013}, who although defining
Ro very differently from our vorticity-based definition here, find 
$\text{Ro} \propto \text{Ro}_\text{p}^{3.84 \pm 0.28}$. We note briefly that they
claim that the value of $\text{Ro}$ is strongly dependent upon the Prandtl number studied, and
that low Ro can be achieved at high Pr without achieving a rotationally constrained flow.
We studied only $\text{Pr} = 1$ here, and leave it to future work to determine if
the scaling of $\text{Ro}_{\text{p}} \propto \text{Pr}^{-1/4}$ is the correct scaling to
predict the evolved Rossby number.

Despite the added complexity of stratification and despite our using stress-free rather than
no-slip boundaries, the boundary layer scaling arguments put forth in \cite{king&all2012} seem
to hold up in our systems. This is reminiscent of what we found in \AB, in which
convection in stratified domains, regardless of Mach number, produced boundary-layer
dominated scaling laws of Nu that were nearly identical to the scaling laws found in
Boussinesq \RB convection.

We close by noting that once \pro$\,$ is chosen such that a convective system has the same
Rossby number as an astrophysical object of choice, it is straightforward to increase the 
turbulent nature of 
simulations by increasing Ra, just as in the non-rotating case.
Although all the results reported here are for a Cartesian geometry with 
antiparallel gravity and rotation, preliminary 3D spherical simulations suggest that 
\pro$\,$ also specifies Ro in more complex geometries (Brown et al. 2019 in prep).

\begin{acknowledgements}
We thank Jon Aurnou and the anonymous referee who independently directed
us to the work of \citet{king&all2012} during the review process, which greatly
helped us understand the results of the experiments here. We further thank
the anonymous referee for other helpful and clarifying suggestions.
This work was supported by NASA Headquarters under the NASA Earth and Space
Science Fellowship Program -- Grant 80NSSC18K1199.
EHA further acknowledges the University of Colorado's George 
Ellery Hale Graduate Student Fellowship.
This work was additionally supported by  NASA LWS grant number NNX16AC92G.  
Computations were conducted 
with support by the NASA High End Computing (HEC) Program through the NASA 
Advanced Supercomputing (NAS) Division at Ames Research Center on Pleiades
with allocation GID s1647.
\end{acknowledgements}

\appendix
\section{Table of Simulations}
\label{appendix:table}
Information for select simulations in this work are shown in Table \ref{table:simulation_info}.
The simulation at minimum (Ra, Ta) and maximum (Ra, Ta) for each of the \pro, \con, and $\mathcal{S}$
paths in Fig.~\ref{fig:parameter_space}b are shown. This information for the displayed simulations
and all other simulations in this work is included as a .csv file in the supplemental materials
and is published online in a Zenodo repository \citep{supp_andersetall2019}.

\begin{deluxetable*}{c c c c c c c c c c c c c c c}
\tabletypesize{\footnotesize}
\caption{Table of simulation information
\label{table:simulation_info}
}
\tablehead{																																															
\colhead{Ra$_{\text{top}}$}	&	\colhead{Ta$_{\text{top}}$}	&	\colhead{Ro$_{\text{p, top}}$}	&	\colhead{Ra$_{\text{mid}}$}	&	\colhead{Ta$_{\text{mid}}$}	&	\colhead{Ro$_{\text{p, mid}}$}	&	\colhead{\con}	&	\colhead{$\mathcal{S}$}	&	\colhead{$L_x/L_z$}	&	\colhead{(nz,	nx,	ny)}	&					\colhead{Ro}      	&	\colhead{Re$_{\parallel}$} 	&	\colhead{Re$_{\perp}$}	&	\colhead{Nu}      	}														
\startdata																																															
\multicolumn{3}{l}{\textbf{Constant \pro$\,$, path III}}\\
1.8	$\times 10^{	5	}$	&	4.1	$\times 10^{	7	}$	&	0.60	&	1.4	$\times 10^{	7	}$	&	3.0	$\times 10^{	9	}$	&	1.03	&	0.067	&	1.1	&	0.51	&	(256,		32,		32)	&	0.015	&	19.4	&	2.5	&	1.2	\\				
1.2	$\times 10^{	9	}$	&	5.2	$\times 10^{	12	}$	&	0.60	&	9.2	$\times 10^{	10	}$	&	3.8	$\times 10^{	14	}$	&	1.03	&	0.015	&	3.0	&	0.07	&	(2048,		64,		64)	&	0.026	&	1771	&	32.0	&	15.4	\\				
5.2	$\times 10^{	2	}$	&	4.6	$\times 10^{	3	}$	&	0.96	&	3.8	$\times 10^{	4	}$	&	3.4	$\times 10^{	5	}$	&	1.64	&	0.333	&	1.13	&	2.28	&	(64,		64,		64)	&	0.074	&	4.2	&	2.5	&	1.2	\\				
3.8	$\times 10^{	8	}$	&	3.1	$\times 10^{	11	}$	&	0.96	&	2.8	$\times 10^{	10	}$	&	2.3	$\times 10^{	13	}$	&	1.64	&	0.035	&	6.0	&	0.12	&	(2048,		64,		64)	&	0.129	&	3906	&	113	&	66.9	\\				
7.9	$\times 10^{	1	}$	&	1.0	$\times 10^{	2	}$	&	1.58	&	5.8	$\times 10^{	3	}$	&	7.4	$\times 10^{	3	}$	&	2.70	&	0.888	&	1.56	&	4.44	&	(64,		64,		64)	&	0.303	&	4.4	&	4.9	&	1.7	\\				
1.4	$\times 10^{	7	}$	&	9.7	$\times 10^{	8	}$	&	1.58	&	1.0	$\times 10^{	9	}$	&	7.2	$\times 10^{	10	}$	&	2.70	&	0.119	&	10.0	&	0.30	&	(512,		128,		128)	&	0.376	&	1257	&	94.9	&	40.1	\\				
\hline																																															
\multicolumn{3}{l}{\textbf{Constant \con$\,$, path II}}\\
8.6	$\times 10^{	4	}$	&	8.6	$\times 10^{	6	}$	&	0.74	&	6.3	$\times 10^{	6	}$	&	6.3	$\times 10^{	8	}$	&	1.26	&	0.1	&	1.47	&	0.68	&	(128,		128,		128)	&	0.051	&	40.2	&	6.7	&	2.2	\\				
2.6	$\times 10^{	6	}$	&	2.6	$\times 10^{	8	}$	&	1.13	&	1.9	$\times 10^{	8	}$	&	1.9	$\times 10^{	10	}$	&	1.93	&	0.1	&	4.64	&	0.39	&	(256,		512,		512)	&	0.27	&	565	&	53.2	&	33.3	\\				
1.4	$\times 10^{	3	}$	&	1.6	$\times 10^{	4	}$	&	1.01	&	1.1	$\times 10^{	5	}$	&	1.2	$\times 10^{	6	}$	&	1.72	&	0.3	&	1.47	&	1.90	&	(64,		128,		128)	&	0.124	&	11.3	&	5.4	&	1.8	\\				
1.1	$\times 10^{	6	}$	&	1.2	$\times 10^{	7	}$	&	2.29	&	7.8	$\times 10^{	7	}$	&	8.6	$\times 10^{	8	}$	&	3.93	&	0.3	&	14.7	&	0.65	&	(192,		384,		384)	&	0.808	&	529	&	83.5	&	27.3	\\				
5.5	$\times 10^{	1	}$	&	5.5	$\times 10^{	1	}$	&	1.65	&	4.0	$\times 10^{	3	}$	&	4.0	$\times 10^{	3	}$	&	2.82	&	1.0	&	1.47	&	4.84	&	(64,		128,		128)	&	0.303	&	3.6	&	4.4	&	1.5	\\				
2.8	$\times 10^{	6	}$	&	2.8	$\times 10^{	6	}$	&	6.39	&	2.0	$\times 10^{	8	}$	&	2.0	$\times 10^{	8	}$	&	10.93	&	1.0	&	100	&	0.82	&	(256,		512,		512)	&	3.357	&	1099	&	220	&	46.6	\\				
\hline																																															
\multicolumn{3}{l}{\textbf{Constant $\mathcal{S}$, path I}}\\
1.9	$\times 10^{	1	}$	&	1.1	$\times 10^{	-1	}$	&	10.00	&	1.4	$\times 10^{	3	}$	&	7.9	$\times 10^{		}$	&	17.12	&	13.2	&	2.0	&	9.91	&	(64,		64,		64)	&	3.668	&	3.0	&	10.2	&	1.7	\\				
3.0	$\times 10^{	6	}$	&	1.1	$\times 10^{	9	}$	&	0.70	&	2.2	$\times 10^{	8	}$	&	8.1	$\times 10^{	10	}$	&	1.20	&	0.052	&	2.0	&	0.30	&	(512,		128,		128)	&	0.053	&	242	&	17.9	&	6.0	\\				
3.0	$\times 10^{	1	}$	&	2.0	$\times 10^{	-1	}$	&	10.00	&	2.2	$\times 10^{	3	}$	&	1.5	$\times 10^{	1	}$	&	17.12	&	12.2	&	3.0	&	9.48	&	(64,		64,		64)	&	4.418	&	5.0	&	15.5	&	2.1	\\				
1.3	$\times 10^{	7	}$	&	5.6	$\times 10^{	9	}$	&	0.80	&	9.7	$\times 10^{	8	}$	&	4.2	$\times 10^{	11	}$	&	1.37	&	0.048	&	3.0	&	0.23	&	(512,		128,		128)	&	0.08	&	592	&	33.4	&	13.6	\\				
\enddata																																															
\tablecomments{
Input parameters and output parameters for select simulations are shown. For each of
the eight paths in Fig. \ref{fig:parameter_space}b, we show information for the lowest and
highest (Ra, Ta) point on that path. The first six rows show information for constant
$\pro\,$ paths, the next six for constant \con$\,$ paths, and the last four for constant $\mathcal{S}$ paths.
We show the input Ra, Ta, and \pro$\,$ at the top of the atmosphere, as well as their
stratification-weighted values at the midplane of the atmosphere, which provide a more direct
comparison to Boussinesq values \citep{unnoetall1960}. We also provide the input \con$\,$ at the
top of the atmosphere, $\mathcal{S}$, aspect ratio ($L_x/L_z$), and coefficient resolution
(nz, nx, ny). Each dimension of the physical grid is 3/2 the size of the coefficient grid 
for adequate dealiasing of quadratic nonlinear terms.
Output values of Ro, Re$_{\parallel}$, Re$_{\perp}$, and Nu are also provided.
This table in its entirety is published as a supplemental \texttt{.csv} file with this
manuscript and also online in a Zenodo repository \citep{supp_andersetall2019}.
}
\end{deluxetable*}


\begin{thebibliography}{}
\expandafter\ifx\csname natexlab\endcsname\relax\def\natexlab#1{#1}\fi

\bibitem[{Ahlers {et~al.}(2009)Ahlers, Grossmann, \& Lohse}]{ahlers&all2009}
Ahlers, G., Grossmann, S., \& Lohse, D. 2009, Rev. Mod. Phys., 81, 503

\bibitem[{{Anders} \& {Brown}(2017)}]{anders&brown2017}
{Anders}, E.~H., \& {Brown}, B.~P. 2017, Physical Review Fluids, 2, 083501

\bibitem[{{Anders} {et~al.}(2019){Anders}, {Manduca}, {Brown}, {Oishi}, \&
  {Vasil}}]{supp_andersetall2019}
{Anders}, E.~H., {Manduca}, C.~M., {Brown}, B.~P., {Oishi}, J.~S., \& {Vasil},
  G.~M. 2019, {Supplemental Materials: Predicting the Rossby number in
  convective experiments}, v.1.0,  Zenodo, doi:10.5281/zenodo.2539500

\bibitem[{{Augustson} {et~al.}(2012){Augustson}, {Brown}, {Brun}, {Miesch}, \&
  {Toomre}}]{augustson&all2012}
{Augustson}, K.~C., {Brown}, B.~P., {Brun}, A.~S., {Miesch}, M.~S., \&
  {Toomre}, J. 2012, \apj, 756, 169

\bibitem[{{Aurnou} \& {King}(2017)}]{aurnou&king2017}
{Aurnou}, J.~M., \& {King}, E.~M. 2017, Proceedings of the Royal Society of
  London Series A, 473, 20160731

\bibitem[{{Brown} {et~al.}(2008){Brown}, {Browning}, {Brun}, {Miesch}, \&
  {Toomre}}]{brown&all2008}
{Brown}, B.~P., {Browning}, M.~K., {Brun}, A.~S., {Miesch}, M.~S., \& {Toomre},
  J. 2008, \apj, 689, 1354

\bibitem[{{Brown} {et~al.}(2010){Brown}, {Browning}, {Brun}, {Miesch}, \&
  {Toomre}}]{brown&all2010}
---. 2010, \apj, 711, 424

\bibitem[{{Brown} {et~al.}(2011){Brown}, {Miesch}, {Browning}, {Brun}, \&
  {Toomre}}]{brown&all2011}
{Brown}, B.~P., {Miesch}, M.~S., {Browning}, M.~K., {Brun}, A.~S., \& {Toomre},
  J. 2011, \apj, 731, 69

\bibitem[{{Brummell} {et~al.}(1996){Brummell}, {Hurlburt}, \&
  {Toomre}}]{brummell&all1996}
{Brummell}, N.~H., {Hurlburt}, N.~E., \& {Toomre}, J. 1996, \apj, 473, 494

\bibitem[{{Brummell} {et~al.}(1998){Brummell}, {Hurlburt}, \&
  {Toomre}}]{brummell&all1998}
---. 1998, \apj, 493, 955

\bibitem[{{Brun} {et~al.}(2017){Brun}, {Strugarek}, {Varela}, {Matt},
  {Augustson}, {Emeriau}, {DoCao}, {Brown}, \& {Toomre}}]{brun&all2017}
{Brun}, A.~S., {Strugarek}, A., {Varela}, J., {et~al.} 2017, \apj, 836, 192

\bibitem[{{Busse}(2002)}]{busse2002}
{Busse}, F.~H. 2002, Physics of Fluids, 14, 1301

\bibitem[{{Calkins} {et~al.}(2015){Calkins}, {Julien}, \&
  {Marti}}]{calkins&all2015a}
{Calkins}, M.~A., {Julien}, K., \& {Marti}, P. 2015, Geophysical and
  Astrophysical Fluid Dynamics, 109, 422

\bibitem[{{Chandrasekhar}(1961)}]{Chandrasekhar}
{Chandrasekhar}, S. 1961, {Hydrodynamic and hydromagnetic stability}

\bibitem[{{Cheng} {et~al.}(2015){Cheng}, {Stellmach}, {Ribeiro}, {Grannan},
  {King}, \& {Aurnou}}]{cheng&all2015}
{Cheng}, J.~S., {Stellmach}, S., {Ribeiro}, A., {et~al.} 2015, Geophysical
  Journal International, 201, 1

\bibitem[{{Ecke} \& {Niemela}(2014)}]{ecke&niemela2014}
{Ecke}, R.~E., \& {Niemela}, J.~J. 2014, \prl, 113, 114301

\bibitem[{{Featherstone} \& {Hindman}(2016)}]{featherstone&hindman2016}
{Featherstone}, N.~A., \& {Hindman}, B.~W. 2016, \apj, 830, L15

\bibitem[{{Gastine} {et~al.}(2016){Gastine}, {Wicht}, \&
  {Aubert}}]{gastine&all2016}
{Gastine}, T., {Wicht}, J., \& {Aubert}, J. 2016, Journal of Fluid Mechanics,
  808, 690

\bibitem[{{Gastine} {et~al.}(2014){Gastine}, {Yadav}, {Morin}, {Reiners}, \&
  {Wicht}}]{gastine&all2014}
{Gastine}, T., {Yadav}, R.~K., {Morin}, J., {Reiners}, A., \& {Wicht}, J. 2014,
  \mnras, 438, L76

\bibitem[{{Glatzmaier} \& {Gilman}(1982)}]{glatzmaier&gilman1982}
{Glatzmaier}, G.~A., \& {Gilman}, P.~A. 1982, \apj, 256, 316

\bibitem[{{Greer} {et~al.}(2016){Greer}, {Hindman}, \&
  {Toomre}}]{greer&all2016}
{Greer}, B.~J., {Hindman}, B.~W., \& {Toomre}, J. 2016, \apj, 824, 4

\bibitem[{{Guerrero} {et~al.}(2013){Guerrero}, {Smolarkiewicz}, {Kosovichev},
  \& {Mansour}}]{guerrero&all2013}
{Guerrero}, G., {Smolarkiewicz}, P.~K., {Kosovichev}, A.~G., \& {Mansour},
  N.~N. 2013, \apj, 779, 176

\bibitem[{{Julien} {et~al.}(2012){Julien}, {Knobloch}, {Rubio}, \&
  {Vasil}}]{julien&all2012}
{Julien}, K., {Knobloch}, E., {Rubio}, A.~M., \& {Vasil}, G.~M. 2012, Physical
  Review Letters, 109, 254503

\bibitem[{{Julien} {et~al.}(1996){Julien}, {Legg}, {McWilliams}, \&
  {Werne}}]{julien&all1996}
{Julien}, K., {Legg}, S., {McWilliams}, J., \& {Werne}, J. 1996, Journal of
  Fluid Mechanics, 322, 243

\bibitem[{{K{\"a}pyl{\"a}} {et~al.}(2014){K{\"a}pyl{\"a}}, {K{\"a}pyl{\"a}}, \&
  {Brandenburg}}]{kapyla&all2014}
{K{\"a}pyl{\"a}}, P.~J., {K{\"a}pyl{\"a}}, M.~J., \& {Brandenburg}, A. 2014,
  \aap, 570, A43

\bibitem[{{King} {et~al.}(2012){King}, {Stellmach}, \& {Aurnou}}]{king&all2012}
{King}, E.~M., {Stellmach}, S., \& {Aurnou}, J.~M. 2012, Journal of Fluid
  Mechanics, 691, 568

\bibitem[{{King} {et~al.}(2013){King}, {Stellmach}, \&
  {Buffett}}]{king&all2013}
{King}, E.~M., {Stellmach}, S., \& {Buffett}, B. 2013, Journal of Fluid
  Mechanics, 717, 449

\bibitem[{{King} {et~al.}(2009){King}, {Stellmach}, {Noir}, {Hansen}, \&
  {Aurnou}}]{king&all2009}
{King}, E.~M., {Stellmach}, S., {Noir}, J., {Hansen}, U., \& {Aurnou}, J.~M.
  2009, \nat, 457, 301

\bibitem[{{Schmitz} \& {Tilgner}(2009)}]{schmitz&tilgner2009}
{Schmitz}, S., \& {Tilgner}, A. 2009, \pre, 80, 015305

\bibitem[{{Soderlund} {et~al.}(2015){Soderlund}, {Sheyko}, {King}, \&
  {Aurnou}}]{soderlund&all2015}
{Soderlund}, K.~M., {Sheyko}, A., {King}, E.~M., \& {Aurnou}, J.~M. 2015,
  Progress in Earth and Planetary Science, 2, 24

\bibitem[{{Stellmach} {et~al.}(2014){Stellmach}, {Lischper}, {Julien}, {Vasil},
  {Cheng}, {Ribeiro}, {King}, \& {Aurnou}}]{stellmach&all2014}
{Stellmach}, S., {Lischper}, M., {Julien}, K., {et~al.} 2014, \prl, 113, 254501

\bibitem[{{Unno} {et~al.}(1960){Unno}, {Kato}, \& {Makita}}]{unnoetall1960}
{Unno}, W., {Kato}, S., \& {Makita}, M. 1960, Publications of the Astronomical
  Society of Japan, 12, 192

\bibitem[{{Zhong} {et~al.}(2009){Zhong}, {Stevens}, {Clercx}, {Verzicco},
  {Lohse}, \& {Ahlers}}]{zhong&all2009}
{Zhong}, J.-Q., {Stevens}, R.~J.~A.~M., {Clercx}, H.~J.~H., {et~al.} 2009,
  Physical Review Letters, 102, 044502

\end{thebibliography}
\end{document}